\documentclass[pra, twocolumn, floatfix, nofootinbib]{revtex4}
\usepackage{graphicx}
\usepackage{color}
\usepackage{amsmath, amsfonts, amssymb, bm}

\begin{document}
\title{Two-center resonant photoionization-excitation\\ driven by combined 
intra- and interatomic electron correlations}
\author{S Kim}
\author{S Steinh\"auser}
\author{A B Voitkiv}
\author{C M\"uller}
\affiliation{Institut f\"ur Theoretische Physik, Heinrich-Heine-Universit\"at D\"usseldorf, Universit\"atsstra{\ss}e 1, 40225 D\"usseldorf, Germany}
\date{\today}
\begin{abstract}
Ionization-excitation of an atom induced by the absorption of a single photon in the presence of a neighbouring atom is studied. The latter is, first, resonantly photoexcited and, afterwards, transfers the excitation energy radiationlessly to the other atom, leading to its ionization with simultaneous excitation. The process relies on the combined effects of interatomic and intraatomic electron correlations. Under suitable conditions, it can dominate by several orders of magnitude over direct photoionization-excitation and even over direct photoionization. In addition, we briefly discuss another kind of two-center resonant photoionization with excitation where the ionization and residual excitation in the final state are located at different atomic sites.  
\end{abstract}

\maketitle

\section{Introduction}

Absorption of a single photon by an atom (or molecule) typically leads to 
excitation or ionization of a single electron. From a theoretical point 
of view this is in line with the fact that photoabsorption is induced by 
a one-body operator. However, due to intraatomic electron correlations, 
the absorption of a single photon may also lead to the simultaneous transition 
of two or more electrons in an atom. This is distinctly exemplified by the 
processes of single-photon double ionization \cite{SPDI-Review1, SPDI-Review2} 
as well as photoionization accompanied by excitation \cite{PIE-Ca-exp, PIE-Ca-1, 
PIE-Ca-2, Salomonson, Burgdoerfer, Gorczyca, Greene, Keifets, Bizau, Wehlitz}, 
which would not exist in the absence of electron correlations. 

Electron correlations also play a prominent role in resonant photoionization.
Here, the absorption of a single photon leads to the population of an 
autoionizing state in the atom, which afterwards deexcites via Auger decay,
releasing an electron into the continuum. 
Resonant photoionization may also occur in two-center atomic systems, where it is 
driven by {\it inter}atomic correlations. Here, an atom $B$ is first resonantly
excited by photoabsorption and, afterwards, deexcites by transfering the excess
energy radiationlessly to a neighboring atom $A$ of different species, causing
its ionization. This two-center resonant photoionization (2CPI) has been studied
both theoretically \cite{2CPI, Perina, 2CPI-coll, 2CPI-mol, ETI} and experimentally 
\cite{2CPIexp, Hergenhahn} in recent years.

We note that the second step in 2CPI -- i.e., the radiationless decay of an excited 
atomic system by energy transfer to a neighboring atomic system upon which the 
latter emits an electron -- is called interatomic Auger decay \cite{Matthew} or 
interatomic Coulombic decay (ICD) \cite{ICD}. The study \cite{ICD} has triggered 
extensive investigations of such decays in various systems such as noble
gas dimers and clusters, both theoretically and experimentally \cite{ICD-Review1, 
ICD-Review2, ICD-Review3}. 

Recent experiments have observed two-electron transitions after single-photon
absorption in dimers and clusters due to interatomic electron correlations.
Photofragmentation of the $^4$He$_2$ dimer into He$^+$ ions, which proceeds via 
photoejection of an electron from one of the helium atoms followed by an ($e$, $2e$) 
reaction at the other atom, was studied experimentally \cite{SPDI-He2e-exp} and 
theoretically \cite{SPDI-He2e-theor}. Collisional mechanisms leading to double 
ionization have also been observed in Ne clusters after resonant inner-valence 
photoexcitation \cite{Hans-2023}. Double ionization of magnesium in Mg-He clusters was found 
to be largely enhanced due to electron-transfer-mediated decay \cite{SPDI-ETMD-theo, 
SPDI-ETMD-exp}: after photoionization, the resulting He$^+$ ion was neutralized 
via electron transfer from a Mg atom and a second electron was simultaneously ejected 
from Mg to keep the energy balance. Double ICD has been predicted to occur in endohedral 
fullerenes such as Mg@C$_{60}$ \cite{DICD-theo, Fedyk}. A related experiment on alkali 
dimers attached to helium droplets observed double ionization followed by dissociation 
of the dimer due to energy transfer from excited helium atoms \cite{DICD-exp}.
Single-photon double ionization due to combined intra- and interatomic correlations 
has been studied theoretically in diatomic systems \cite{Eckey2020}. 

\begin{figure}[b]
\begin{center}
\includegraphics[width=0.4\textwidth]{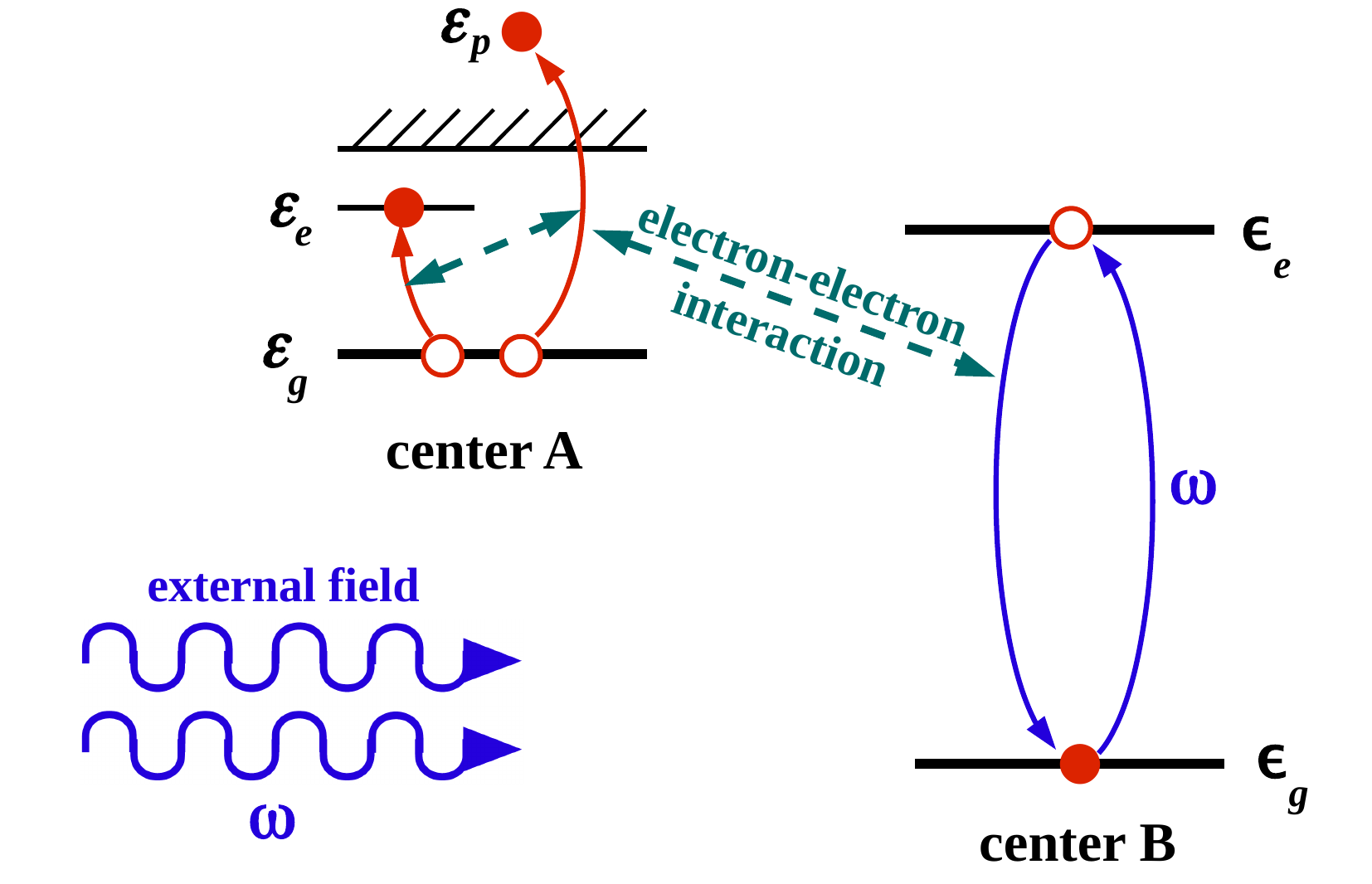}
\caption{Scheme of two-center resonant photoionization-excitation (2CPIE). First, atom $B$ is resonantly photoexcited. Afterwards, upon radiationless energy transfer to atom $A$, the latter is singly ionized and simultaneously excited. The process involves three active electrons and relies on both interatomic and intraatomic electron correlations, as indicated by the dashed arrows.}
\label{fig:scheme}
\end{center}
\end{figure}

In the present paper, we study resonant photoionization with simultaneous excitation 
in two-center atomic systems. The process relies on the joint action of intra\-atomic and 
interatomic electron correlations. In a first step, an atom $B$ is resonantly excited
by photoabsorption. The excitation energy is sufficiently large to lead -- upon 
radiationless energy transfer to a neighbouring atom $A$ of different atomic species --
to ionization-excitation of the latter (see Fig.~\ref{fig:scheme}). The correlation-driven 
process may be termed two-center resonant photoionization-excitation (2CPIE). We will show 
that, under suitable conditions, the cross section for 2CPIE can largely exceed the cross sections
for direct photoionization-excitation and even for direct photoionization of an isolated 
atom $A$. In addition, we will discuss a variant of 2CPI where atom $B$ deexcites only 
partially, leading to ionization of atom $A$, while atom $B$ ends up in an excited state.

From photoionization studies of single atoms it is known that photoionization-excitation 
is a relevant process which leads to satellite lines in the photoionization cross section 
\cite{PIE-Ca-1}. The cross section for photoionization with excitation can be comparable 
or even larger than the cross section for single-photon double ionization \cite{Bizau}.

Our paper is organized as follows. In Sec.~II we present our theoretical approach to 
2CPIE and discuss its relation with one-center photoionization-excitation. In Sec.~III 
we apply our formalism to various two-center atomic systems and demonstrate 
the relevance of 2CPIE. In Sec.~IV the related process of 2CPI of an atom $A$ with residual 
excitation in a neighbour atom $B$ is introduced and discussed. Conclusions are given in Sec.~V. 

Atomic units (a.u.) are used throughout unless explicitly stated otherwise.

\section{Theory of two-center resonant photoionization-excitation}
\subsection{General considerations}

Let us consider a system consisting of two atoms, $A$ and $B$, separated 
by a sufficiently large distance $R$ such that their individuality 
is basically preserved. The atoms, which are initially in their ground states,  
are exposed to a resonant electromagnetic field. The latter 
will be treated as a classical electromagnetic wave of linear polarization, 
whose electric field component reads 
\begin{eqnarray}
{\bf F}({\bf r},t)= {\bf F}_0 \cos\left(\omega t - {\bf k} \cdot {\bf r}\right).
\label{field}
\end{eqnarray}
Here $\omega = c k $ and ${\bf k}$ 
are the angular frequency and wave vector,  
and ${\bf F}_0=F_0\,{\bf e}_z$ denotes the field strength vector
which is chosen to define the $z$ direction.

Assuming the atoms to be at rest,  
we take the position of the nucleus of atom $A$  
as the origin and denote the coordinates 
of the nucleus of atom $B$, the two (active) electrons of atom $A$ 
and that of atom $B$ by ${\bf R}$, ${\bf r}_j$ ($j\in\{1,2\}$) 
and ${\bf r}_3 ={\bf R} + \boldsymbol{\xi}$, 
respectively, where $\boldsymbol{\xi}$ 
is the position of the electron of atom $B$ 
with respect to its nucleus. 
Let atom $B$ have an excited state $\chi_e$
reachable from the ground state $\chi_g$ by a dipole-allowed transition. 

The total Hamiltonian describing the
two atoms in the external electromagnetic field reads 
\begin{eqnarray} 
H = \hat{H}_0 + \hat{V}_{AB} + \hat{W},  
\label{hamiltonian}  
\end{eqnarray} 
where $ \hat{H}_0 $ is the sum of the Hamiltonians 
for the noninteracting atoms $A$ and $B$,  
$\hat{V}_{AB}$ the interaction between the atoms 
and $\hat{W} = \hat{W}_A + \hat{W}_B$ the interaction 
of the atoms with the electromagnetic field.
Within the dipole approximation and length gauge, the interaction $\hat{W}$ reads
\begin{eqnarray} 
\hat{W} = \sum_{j=1,2,3} {\bf F}({\bf r}_j={\bf 0},t) \cdot {\bf r}_j\ .
\label{W} 
\end{eqnarray} 
The two terms with $j\in\{1,2\}$ compose the interaction $\hat{W}_A$ with the electrons in atom $A$,
whereas the term with $j=3$ the interaction $\hat{W}_B$ with the electron in atom $B$.
For electrons undergoing electric dipole transitions, the interatomic 
interaction reads 
\begin{eqnarray} 
\hat{V}_{AB} = \sum_{j=1,2}\left( \frac{{\bf r}_j\cdot \boldsymbol{\xi}}{R^3} 
- \frac{ 3 ({\bf r}_j\cdot{\bf R})(\boldsymbol{\xi}\cdot{\bf R})}{R^5} \right)\ .
\label{V_AB} 
\end{eqnarray} 
It is assumed that $\omega_{ge} R /c  \ll 1$, where $\omega_{ge} = \epsilon_e - \epsilon_g$ 
is the atomic transition frequency and $c$ the speed of light, such that 
retardation effects can be neglected.

In the process of 2CPIE one has essentially three different basic three-electron configurations, which are schematically illustrated in Fig.~\ref{fig:scheme}:
(I) $\Psi_{g;g} = \Phi_{g}({\bf r}_1, {\bf r}_2) \chi_g(\boldsymbol{\xi})$ with total energy $E_{g;g} = \varepsilon_g + \epsilon_g$, where both atoms are in their corresponding ground states $\Phi_g$ and $\chi_g$;
(II) $\Psi_{g;e} = \Phi_{g}({\bf r}_1, {\bf r}_2) \chi_e(\boldsymbol{\xi})$ with total energy $E_{g;e} = \varepsilon_g + \epsilon_e$, in which atom $A$ is in the ground state while atom $B$ is in the excited state $\chi_e$;
(III) $\Psi_{{\bf p},e;g} = \Phi_{{\bf p},e}({\bf r}_1,{\bf r}_2) \chi_g(\boldsymbol{\xi})$ with total energy $E_{{\bf p}, e; g} = \varepsilon_{{\bf p},e} + \epsilon_g$ and $\varepsilon_{{\bf p},e} = \varepsilon_p + \varepsilon_e$, where one electron of atom $A$ has been emitted into the continuum with asymptotic momentum ${\bf p}$ and energy $\varepsilon_p=\frac{p^2}{2}$, whereas the other one has been excited to a bound state of energy $\varepsilon_e$, while the electron of atom $B$ has returned to the ground state. 

Within the second order of time-dependent perturbation theory, the probability amplitude for 2CPIE can be written as
\begin{eqnarray}
S^{(2)}_{{\bf p},e}\! &=&\! -\int_{-\infty}^{\infty} dt\, \langle \Psi_{{\bf p},e;g}|\hat{V}_{AB}| \Psi_{g;e} \rangle\, e^{-i(E_{g;e}-E_{{\bf p},e;g})t}\nonumber\\
& & \times \int_{-\infty}^{t} dt'\, \langle \Psi_{g;e}|\hat{W}_{B}| \Psi_{g;g} \rangle\, e^{-i(E_{g;g}-E_{g;e})t'}\ .
\label{S1}
\end{eqnarray}

By performing the inner time integral, we obtain
\begin{eqnarray} 
S^{(2)}_{{\bf p},e} &=& -i\int_{-\infty}^{\infty} dt\, \langle \Psi_{{\bf p},e;g}|\hat{V}_{AB}| \Psi_{g;e} \rangle\,\frac{F_0}{2} \nonumber\\
& & \times\,\frac{\langle \chi_e|\xi_z| \chi_g \rangle}{\omega-\omega_{ge}+\frac{i}{2}\Gamma}\,e^{-i(E_{g;g}+\omega-E_{{\bf p},e;g})t}\ .
\label{S2}
\end{eqnarray}
Here we have kept only the term with the resonant denominator and inserted the total 
width $\Gamma=\Gamma_{\rm rad}+\Gamma_{_{\rm ICD}}$ of the excited state $\chi_e$ in 
atom $B$. It accounts for the finite lifetime of this state and consists of the radiative width 
\begin{eqnarray}
\Gamma_{\rm rad} = \frac{4\omega_{ge}^3}{3c^3}\big| \langle \chi_e|\boldsymbol{\xi}|\chi_g\rangle \big|^2
\label{Gamma_rad}
\end{eqnarray}
and the ICD width
\begin{eqnarray}
\Gamma_{_{\rm ICD}} = \frac{p'}{(2\pi)^2}\int{\rm d}\Omega_{{\bf p}'} \big|\langle\Phi_{{\bf p}',g}\chi_g|\hat{V}_{AB}|\Phi_g\chi_e\rangle\big|^2\ ,
\label{Gamma_ICD}
\end{eqnarray}
where the integral is taken over the emission angles of the ICD electron that is ejected from atom $A$,
and $\Phi_{{\bf p}',g}$ denotes the state of atom $A$ where one electron has been emitted with asymptotic momentum ${\bf p}'$ and energy $\varepsilon_{p'}=\varepsilon_g + \omega_{ge}-\varepsilon_g^+$, while the other electron is in the ground state with energy $\varepsilon_g^+$ of the resulting singly charged ion.
We note that, in the considered scenario, there is an additional contribution to $\Gamma_{_{\rm ICD}}$ associated with the decay channel where  deexcitation of atom $B$ leads to photoionization with excitation of atom $A$. This contribution, however, is usually small and may be neglected.

Taking also the outer time integral, we arrive at
\begin{eqnarray} 
S^{(2)}_{{\bf p},e} &=& -i\pi\, \langle \Phi_{{\bf p},e}|({\bf r}_1 + {\bf r}_2)| \Phi_g \rangle\cdot\left( {\bf e}_z - \frac{3R_z}{R^2}\,{\bf R}\right) \nonumber\\
& & \times\ \frac{F_0}{R^3}\, \frac{\left| \langle \chi_e|\xi_z| \chi_g \rangle \right|^2}{\Delta+\frac{i}{2}\Gamma}\ \delta(\varepsilon_{p} + \varepsilon_e - \varepsilon_g - \omega)\, ,
\label{S3}
\end{eqnarray}
where the detuning from the resonance $\Delta = \omega-\omega_{ge}$ has been introduced.
The delta function in Eq.~\eqref{S3} displays the energy conservation in the process.
In this relation, the energies of atom $B$ have dropped out, in accordance with its role as catalyzer.

From the transition amplitude we can obtain the fully differential ionization cross section in the usual way by taking the absolute square and dividing it by the interaction time $\tau$ and the incident flux $j=\frac{cF_0^2}{8\pi \omega}$, that is
\begin{eqnarray} 
\frac{{\rm d}^3\sigma_{_{\rm 2CPIE}}}{{\rm d}^3p} = \frac{1}{(2\pi)^3 j\tau}\,\left|S^{(2)}_{{\bf p},e}\right|^2\ .
\label{CS}
\end{eqnarray} 
The factor $(2\pi)^{-3}$ arises from the fact that the continuum states in our calculations are normalized to a quantization volume of unity. Performing the integration over $\varepsilon_{p}$ with the help of the $\delta$-function in Eq.~\eqref{S3}, we obtain the angle-differential cross section
\begin{eqnarray} 
\frac{{\rm d}\sigma_{_{\rm 2CPIE}}}{{\rm d}\Omega_{p}} &=& \frac{\omega\,p_c}{(2\pi)^2 c R^6}\ \frac{\left| \langle \chi_e|\xi_z| \chi_g \rangle \right|^4}{\Delta^2+\frac{1}{4}\Gamma^2}\nonumber\\
& & \hspace{-1.5cm} \times\, \big| \langle \Phi_{{\bf p},e}|({\bf r}_1 + {\bf r}_2)| \Phi_g \rangle \cdot\left( {\bf e}_z - 3\cos\theta_R\,{\bf e}_R\right) \big|^2_{p=p_c}
\label{CS1}
\end{eqnarray}
where we have introduced the unit vector ${\bf e}_R={\bf R}/R$ along the internuclear separation and the angle $\theta_R$ between ${\bf R}$ and the field direction. Besides, $p_c$ denotes the momentum value of the electron emitted into the continuum, as determined by the $\delta$-function in Eq.~\eqref{S3}.

\subsection{Relation to one-center processes}

We can draw a comparison with the direct photo\-ionization-excitation of atom $A$ by the electromagnetic field. The corresponding probability amplitude in the first order of perturbation theory is given by 
\begin{eqnarray}
S^{(1)}_{{\bf p},e} &=& -i\int_{-\infty}^{\infty} dt\, \langle \Phi_{{\bf p},e}|\hat{W}_A| \Phi_g \rangle\, e^{-i(\varepsilon_g - \varepsilon_{p} - \varepsilon_e)t}\nonumber\\
&=& -i\,\frac{F_0}{2}\, \langle \Phi_{{\bf p},e}|({\bf r}_1 + {\bf r}_2)\cdot{\bf e}_z| \Phi_g \rangle \nonumber\\
& & \times\ 2\pi\delta(\varepsilon_{p} + \varepsilon_e - \varepsilon_g - \omega)\ .
\label{S-1C}
\end{eqnarray}
For the special cases, when the separation vector ${\bf R}$ between the atoms $A$ and $B$ is oriented either along the field direction or perpendicular to it, we can cast Eq.\,\eqref{S3} into the form
\begin{eqnarray} 
S^{(2)}_{{\bf p},e} \,=\, \frac{\alpha}{R^3}\,\frac{\left| \langle \chi_e|\xi_z| \chi_g \rangle \right|^2}{\Delta+\frac{i}{2}\Gamma}\,S^{(1)}_{{\bf p},e}
\label{S4}
\end{eqnarray}
which translates into the relation 
\begin{eqnarray} 
\sigma_{_{\rm 2CPIE}} = \frac{\alpha^2}{R^6}\,\frac{\left| \langle \chi_e|\xi_z| \chi_g \rangle \right|^4}{\Delta^2+\frac{1}{4}\Gamma^2}\,\sigma_{_{\rm PIE}}
\label{CS2}
\end{eqnarray}
between the corresponding cross sections. Here, $\alpha=-2$ for ${\bf R}\parallel {\bf F}_0$ and $\alpha=1$ for ${\bf R}\perp {\bf F}_0$. 
When the field is exactly resonant with the transition $\chi_g\to\chi_e$ in atom $B$ and the interatomic distance is sufficiently large, so that $\Gamma_{_{\rm ICD}}\ll\Gamma_{\rm rad}$, this expression becomes 
\begin{eqnarray} 
\sigma_{_{\rm 2CPIE}} = \frac{9\alpha^2}{4} 
\left(\frac{c}{\omega R}\right)^6\sigma_{_{\rm PIE}}\ .
\label{CS3}
\end{eqnarray}
Here we have used formula~\eqref{Gamma_rad} for the radiative width. 

Eq.\,\eqref{CS3} shows that the cross section for 2CPIE can be largely enhanced by a factor $[c/(\omega R)]^6\gg 1$ as compared with the usual one-center process of photoionization-excitation, where the neighboring atom $B$ is not involved. For example, assuming $\omega\approx 10$\,eV (corresponding to the first excitation energy in hydrogen) and $R=10$\,\AA, an enormous enhancement by 8 orders of magnitude results. This implies further that 2CPIE can even very strongly exceed the direct photoionization of atom $A$. Because the ratio between photoionization-excitation and photoionization (without excitation) is typically of the order of few percent \cite{Burgdoerfer, Gorczyca, Greene, Keifets, Bizau, Wehlitz}, an enhancement by six orders remains. We point out that, nevertheless, 2CPIE is usually not the dominant ionization channel because 2CPI is much stronger, being enhanced over one-center photoionization by a factor $[c/(\omega R)]^6$ as well \cite{2CPI}. The ratio of 2CPIE--to--2CPI is therefore of similar size as the ratio between the corresponding one-center processes. In special cases, however, 2CPIE can be comparable or even larger than 2CPI (just as single-center PIE can sometimes dominate over single-center photoionization \cite{PIE-Ca-1}); see Sec.~III.

The processes of 2CPIE and direct one-center photoionization-excitation lead to the same final state, since atom $B$ eventually returns to its ground state and thus serves as a catalyzer. Therefore, the corresponding probability amplitudes  \eqref{S3} and \eqref{S-1C} are generally subject to quantum interference. However, for parameters where the two-center channel strongly dominates, the interference is of minor importance and may be neglected.

The extremely high efficiency of 2CPIE arises from its resonant nature, whereas  
one-center photoionization-excitation is a nonresonant process, in general. We note, 
however, that under special circumstances also one-center PIE can proceed in a resonant way: 
photoabsorption by an atom may lead to the population of an intra\-atomic autoionizing 
state, which is able to decay, upon electron emission, not only to the ground state but 
also to an excited state of the resulting ion. In the latter case, the resonant 
photoexcitation of the autoionizing state leads to photoionization-excitation in an 
isolated atom. Corresponding resonance structures in PIE have been predicted to occur 
in He between about 70\,eV to 73\,eV, stemming from doubly excited states \cite{Salomonson}. 
Another example is PIE of Ca through excitation of the $3p\to 3d$ resonance at about 
31\,eV \cite{PIE-Ca-1}.

\section{Numerical Examples and Discussion}
In this section, we illustrate characteristic properties of 2CPIE by way of some concrete examples.

We first consider a two-center system composed of a neutral He atom as center $A$ and a Li$^+$ ion as center $B$. Our consideration is motivated by the fact that helium constitutes a benchmark for photoionization-excitation studies \cite{Salomonson, Burgdoerfer, Gorczyca, Greene, Keifets, Bizau, Wehlitz}. The threshold energy amounts to 65.4\,eV, when the electron in the created He$^+$ ion occupies an $n=2$ state. If a singly charged Li$^+$ ion is located in close vicinity to a He atom and is subject to an external field resonant to its $1s^2\to 1s3p$ transition at $\omega_{ge}\approx 69.65$\,eV \cite{NIST}, photoionization-excitation of helium via 2CPIE may occur. Its cross section is given by Eq.~\eqref{CS2} as
\begin{eqnarray} 
\sigma_{_{\rm 2CPIE}} = \left(\frac{3c^3}{2 \omega_{ge}^3 R^3}\right)^{\!2}
\frac{\Gamma_{\rm rad}^2}{\Delta^2+\frac{1}{4}\Gamma^2}\,\sigma_{_{\rm PIE}}\ ,
\label{CS4}
\end{eqnarray}
assuming for definiteness that the internuclear axis lies along the field direction.
The radiative and ICD decay widths are given by $\Gamma_{\rm rad} \approx 7.76\times 10^9$\,s$^{-1}$
\cite{NIST} and $\Gamma_{_{\rm ICD}}\approx\frac{3c^4}{2\pi\omega_{ge}^4 R^6}\,\Gamma_{\rm rad}\,\sigma_{_{\rm PI}}(\omega_{ge})$ (see, e.g., \cite{ICD-formula}), where $\sigma_{_{\rm PI}}(\omega_{ge})\approx 1$\,Mb \cite{Bizau} denotes the single-center photoionization cross section of helium. Accordingly, $\Gamma_{_{\rm ICD}}/\Gamma_{\rm rad} \approx (7.2/R\,[{\rm a.u.}])^6$. If the resonance condition is exactly met ($\Delta = 0$) and $\Gamma_{\rm rad}\gg\Gamma_{_{\rm ICD}}$, the enhancement of 2CPIE over one-center PIE amounts to $\sigma_{_{\rm 2CPIE}}/\sigma_{_{\rm PIE}}\sim 2\times 10^{11}/(R\,[{\rm a.u.}])^6$. For instance, at $R=10\,$a.u., the 2CPI cross section is larger than $\sigma_{_{\rm PIE}}\approx 0.1$\,Mb \cite{Bizau, Wehlitz} by five orders of magnitude.
Due to of this huge difference, the quantum interference that both processes 
are subject to, is immaterial for the considered parameters.

\medskip

The very large enhancement results from our assumption of exact resonance. In an experiment,
however, the applied photon beam will not be perfectly monochromatic but have a certain frequency 
width $\Delta\omega$. For a typical value $\Delta\omega \sim 1$\,meV (see, e.g., \cite{2CPIexp})
this beam width is much larger than $\Gamma_{\rm rad}\approx 5\,\mu$eV in our example.
While the cross section for nonresonant single-center photoionization-excitation would 
be practically constant over the bandwidth of the photon beam, 2CPIE would only proceed
efficiently for those beam frequencies that are very close to the resonant value 
$\omega_{ge}$. As a consequence, if a He-Li$^+$ system is exposed to a photon beam 
whose frequency range encompasses the resonant frequency $\omega_{ge}$, the ratio 
of cross sections given above will be reduced approximately by a factor 
$\Gamma_{\rm rad}/\Delta\omega\sim 5\times 10^{-3}$. At $R = 10$\,a.u., 
the beam-averaged cross section for 2CPIE will still exceed the one-center 
cross section $\sigma_{_{\rm PIE}}$ by three orders of magnitude.

Alternative photon energies to induce 2CPIE in a He-Li$^+$ system would be
$\omega \approx 72.26$\,eV and $\omega\approx 73.48$\,eV, which are resonant to the 
$1s^2\to 1s4p$ and $1s^2\to 1s5p$ dipole transitions in Li$^+$, respectively \cite{NIST}.

\medskip

Apart from He-Li$^+$, there is also a number of conceivable systems 
composed of two neutral atoms, wherein 2CPIE can proceed. As relation \eqref{CS3} shows, 
it is advantageous for 2CPIE if low interatomic energy transfers are sufficient to induce 
the process in atom $A$. Table I lists a few examples of atoms with rather low 
ionization-excitation energies, which would represent suitable candidates.

 \begin{center}
  \begin{table}[h]
   \begin{tabular}{l||c|c|c|c}
    \hline
    \hline
    Atom $A$ & C & O & Mg & Ca \\
    \hline
    \hline    
    First ionization energy & 11.26\,eV & 13.62\,eV & 7.65\,eV & 6.11\,eV\\                                    
    \hline                 
    Excitation energy of $A^+$ & 9.29\,eV & 3.32\,eV & 4.42\,eV & 3.12\,eV\\
    \hline
    \hline
   \end{tabular}
   \caption{Examples of candidates of atoms $A$ and 
   transition energies in $A^+$, which are suitable for the 2CPIE process.}
  \end{table}
 \end{center}

\vspace{-0.25cm}

(i) As a first diatomic system, let us consider the combination of Ca as atom $A$ with 
H as atom $B$. By an external field of frequency $\omega\approx 10.2$\,eV the $1s\to 2p$ 
transition in H is resonantly excited, whose energy can be transfered to a neighbouring
Ca($4s^2$), causing its ionization with $4s\to 4p$ excitation. The enhancement of 2CPIE
over single-center photoionization-excitation would be of order $\sim (365/R\,{\rm [a.u.]})^6$
which considerably exceeds unity up to interatomic distances of several nanometers. Also
in absolute terms the 2CPIE cross section can be very large, given that 
$\sigma_{_{\rm PIE}}\sim 0.1$\,Mb at $\omega\approx 10$\,eV \cite{PIE-Ca-1,PIE-Ca-2}.

Interestingly, 2CPIE could even compete with 2CPI in this case, because for an 
isolated Ca atom at this energy one has $\sigma_{_{\rm PIE}}\gtrsim\sigma_{_{\rm PI}}$ 
\cite{PIE-Ca-1}, since $\sigma_{_{\rm PI}}$ runs through a Cooper minimum there \cite{ICDexc}.

(ii) A diatomic system with very similar properties would be Mg as atom $A$ and H as atom $B$,  
where the latter is resonantly excited from $1s\to 3p$ at $\omega\approx 12.09$\,eV, 
which can be transfered to Mg($3s^2$), causing its ionization with $3s\to 3p$ excitation
very close to threshold.

(iii) Finally, one can imagine any of the atoms listed in Table I in the vicinity of He as
atom $B$. This could be realized, for instance, by attachment to He droplets, similarly to 
the recent experiments \cite{SPDI-ETMD-exp,DICD-exp}. Exposing the system to an external
field of frequency $\omega\approx 21.2$\,eV, the $1s^2\to 1s2p$ transition in He would be
resonantly excited, with subsequent interatomic energy transfer to, e.g, Mg($3s^2$), causing 
its ionization with simultaneous $3s\to 3p$ (at 4.42\,eV) or $3s\to 4s$ (at 8.65\,eV) or 
$3s\to 4p$ (at 10.0\,eV) excitation of the Mg$^+$ ion. Here, the respective enhancements 
of 2CPIE over single-center PIE would amount to $\sim (176/R\,{\rm [a.u.]})^6$.

It is interesting to note that an early PIE experiment relied on a rather similar setup. 
In \cite{PIE-Ca-exp}, the resonance radiation from He at 21.2\,eV was used to study 
photoionization-excitation of Ca atoms.

\section{2CPI with residual excitation}
\begin{figure}[h]
\begin{center}
\includegraphics[width=0.4\textwidth]{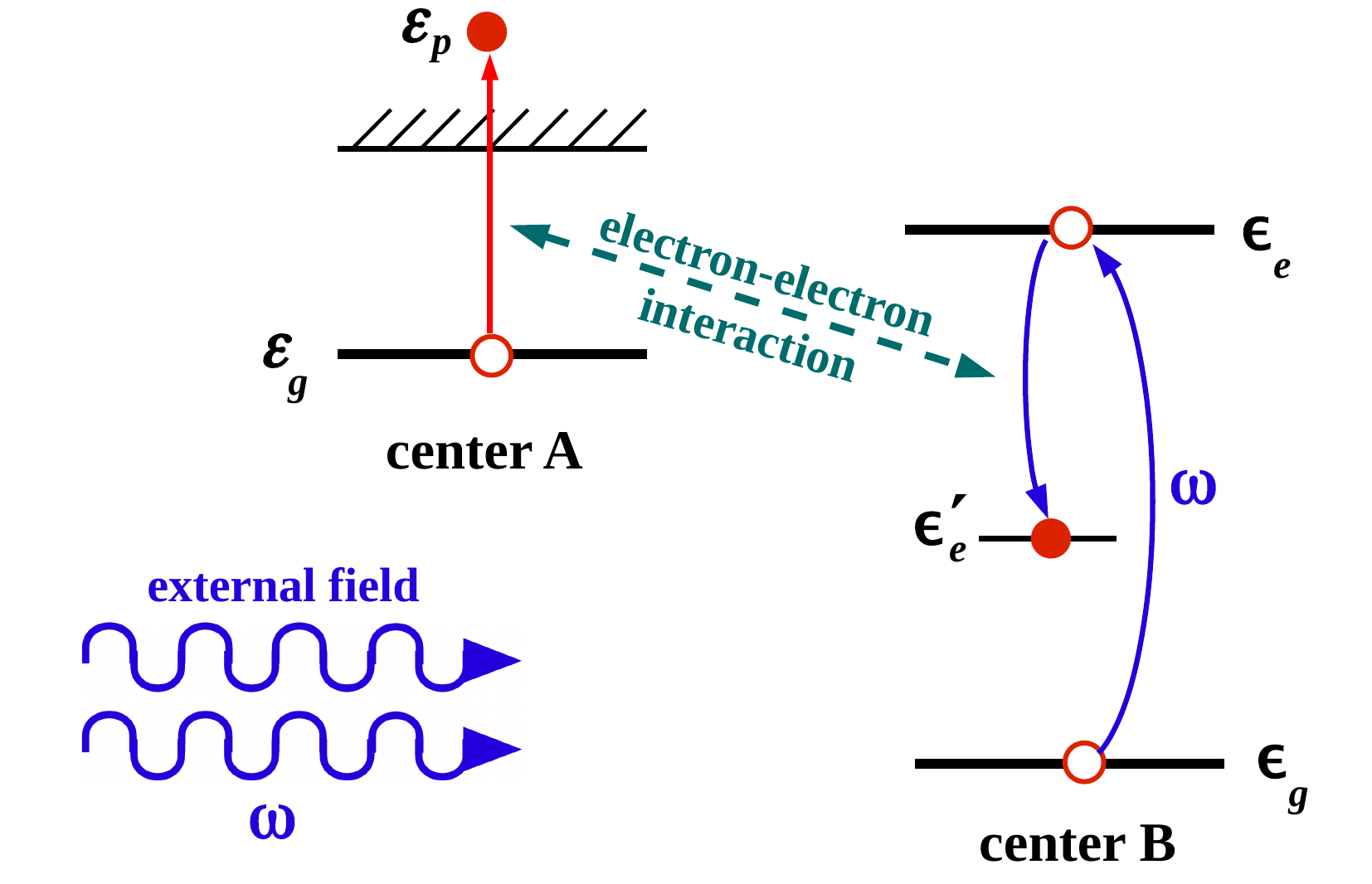}
\caption{2CPI with residual excitation. As in Fig.~\ref{fig:scheme}, atom $B$ is first resonantly photoexcited, but afterwards deexcites not fully down to its initial, but to an intermediate excited state whose energy $\epsilon_e'$ lies in between $\epsilon_g$ and $\epsilon_e$. The energy set free still suffices to singly ionize the neighbouring atom $A$. The process involves two active electrons and relies solely on interatomic electron correlations.}
\label{fig:scheme2}
\end{center}
\end{figure}

Before moving on to the conclusion we briefly describe a variant of 2CPI that is related to 2CPIE. In the standard form of 2CPI, an atom $B$ is resonantly photoexcited and, afterwards, deexcites back into its initial state (typically the ground state), transfering the energy release to a neighbouring atom $A$ which gets ionized \cite{2CPI, Perina, 2CPI-coll, 2CPI-mol}. However, as Fig.~\ref{fig:scheme2} shows, it is also conceivable that the deexcitation of atom $B$ does not proceed fully down to the initial state, but instead to another excited state of intermediate energy. If the corresponding energy difference $\epsilon_e-\epsilon'_e$ is larger than the ionization potential of atom $A$, then the latter could still be ionized via interatomic energy transfer from atom $B$. Accordingly, in this process -- which could be termed 2CPI with residual excitation -- the two-center system ends up in an ionized-excited state, but in contrast to 2CPIE the ionization and excitation are located at two different centers. 

The probability amplitude for 2CPI with residual excitation is given by
\begin{eqnarray}
S^{(2)}_{{\bf p};e'}\! &=&\! -\int_{-\infty}^{\infty} dt\, \langle \Psi_{{\bf p};e'}|\hat{V}_{AB}| \Psi_{g;e} \rangle\, e^{-i(E_{g;e}-E_{{\bf p};e'})t}\nonumber\\
& & \times \int_{-\infty}^{t} dt'\, \langle \Psi_{g;e}|\hat{W}_{B}| \Psi_{g;g} \rangle\, e^{-i(E_{g;g}-E_{g;e})t'}
\label{S1-2CPIe}
\end{eqnarray}
Initial state $\Psi_{g;g}$ and intermediate state $\Psi_{g;e}$ are identical to 2CPI, but the final state is $\Psi_{{\bf p};e'} = \Phi_{{\bf p}}({\bf r}) \chi_e'(\boldsymbol{\xi})$ with total energy $E_{{\bf p}; e'} = \varepsilon_{p} + \epsilon_e'$. The process neither interferes with direct photoionization of atom $A$ nor 2CPI of atoms $A$ and $B$, since it leads to a different final state. 

Similarly to Eq.~\eqref{CS2}, one can express the cross section 
for 2CPI with residual excitation as
\begin{eqnarray}
\sigma^{(e')}_{_{\rm 2CPI}} = \frac{\alpha^2}{R^6}\,\frac{\left| \langle \chi_e'|\xi_z| \chi_e \rangle\,
\langle \chi_e|\xi_z| \chi_g \rangle \right|^2}{\Delta^2+\frac{1}{4}\Gamma^2}\,
\frac{\omega_{ge}^2}{\omega_{e'e}^2}\,\sigma_{_{\rm PI}}(\omega_{e'e})
\label{CSexc}
\end{eqnarray}
where the relation $\omega_{e'e}\approx\omega-\omega_{ge'}$, with 
$\omega_{e'e} = \epsilon_e-\epsilon_e'$ and $\omega_{ge'} = \epsilon_e'-\epsilon_g$, 
has been used. The ratio with ordinary 2CPI (without residual excitation) thus becomes
\begin{eqnarray}
\frac{\sigma^{(e')}_{_{\rm 2CPI}}}{\sigma_{_{\rm 2CPI}}} = \frac{\Gamma_{\rm rad}^{(e'e)}}{\Gamma_{\rm rad}^{(ge)}}\left(\frac{\omega_{ge}}{\omega_{e'e}}\right)^{\!5} \frac{\sigma_{_{\rm PI}}(\omega_{e'e})}{\sigma_{_{\rm PI}}(\omega_{ge})}\ .
\label{CSratio}
\end{eqnarray}
Here, $\Gamma_{\rm rad}^{(e'e)}$ and $\Gamma_{\rm rad}^{(ge)}$ denote the rates 
for the radiative decays $\chi_e\to\chi_e'$ and $\chi_e\to\chi_g$, respectively.
Note that the cross section for ordinary 2CPI \cite{2CPI} can be obtained from 
Eq.~\eqref{CSexc} by formally setting $\chi_e' \equiv \chi_g$ and 
$\epsilon_e'\equiv\epsilon_g$ therein.

Under suitable conditions, 2CPI with residual excitation can have a larger cross section than usual 2CPI \cite{Sebastian}. This is because smaller energy transfers are generally beneficial for the efficiency of interatomic processes, which can overcompensate a possibly smaller transition matrix element (or decay rate) between the states $\chi_e$ and $\chi_e'$ in atom $B$ as compared with $\chi_e$ and $\chi_g$. 

As an example, let us consider a system composed of a neutral Ca atom 
as center $A$ and a singly charged Li$^+$ ion as center $B$. The system is irradiated
by an electromagnetic field of frequency $\omega\approx 69.65$\,eV, resonant with the
$1s^2\to 1s3p$ transition in Li$^+$. After resonant photoexcitation, the $1s3p$ state 
may either decay back into the ground state or, alternatively, into the intermediate
$1s2s$ excited state, releasing an energy of $\omega_{e'e}\approx 8.73$\,eV. In the
first case, 2CPI may happen, whereas in the second case, 2CPI with residual excitation
can arise. According to Eq.~\eqref{CSratio}, the ratio between the corresponding cross
sections amounts to $\sigma^{(e')}_{_{\rm 2CPI}}/\sigma_{_{\rm 2CPI}}\sim 10^3$, where we 
have used the decay rates $\Gamma_{\rm rad}^{(e'e)}\approx 2.83 \times 10^8$\,s$^{-1}$, 
$\Gamma_{\rm rad}^{(ge)}\approx 7.76 \times 10^9$\,s$^{-1}$ \cite{NIST} and 
$\sigma_{_{\rm PI}}(\omega_{e'e})\gtrsim\sigma_{_{\rm PI}}(\omega_{ge})\sim 0.1$\,Mb 
\cite{PIE-Ca-1}. Thus, for the considered Ca-Li$^+$ system and resonant field frequency, 
2CPI with residual excitation can strongly outperform ordinary 2CPI.

We finally note that the $1s2s$ state reached in Li$^+$ after 2CPI with residual
excitation is metastable. The energy stored in this state might therefore induce 
subsequent reactions, such as Penning ionization, for example.

\section{Conclusion}
The process of two-center resonant photoionization-excitation has been studied where the resonant photoexcitation of an atom $B$ leads to ionization with simultaneous excitation of a neighbouring atom $A$ via the combined action of interatomic and intra-atomic electron–electron correlations. It was shown that, due to its resonant character, 2CPIE can largely dominate over the ordinary one-center photoionization-excitation of an isolated atom $A$. The enhancement may persist for interatomic distances up to a several nanometers, as was demonstrated by considering various two-center systems. 2CPIE could in principle be observed in an experimental setup similar to those in \cite{SPDI-ETMD-exp, DICD-exp}.

In addition, we briefly discussed the related process of 2CPI with residual excitation, where the excitation energy of a resonantly photoexcited atom $B$ is only partially transfered to a neighbour atom $A$, leading to its ionization, while the rest of the energy remains in atom $B$. It was shown that this partial interatomic energy transfer may lead to very efficient ionization. Under suitable conditions, it can largely dominate over the ordinary 2CPI where the full excitation energy is transfered.



\begin{thebibliography}{33}

\bibitem{SPDI-Review1} Briggs J S and Schmidt V 2000 {\it J. Phys. B: At. Mol. Opt. Phys.} \textbf{33} R1

\bibitem{SPDI-Review2} Avaldi L and Huetz A 2005 {\it J. Phys. B: At. Mol. Opt. Phys.} \textbf{38} S861

\bibitem{PIE-Ca-exp} Suzer S, Lee S T, and Shirley D A 1976 {\it Phys. Rev. A} {\bf 13} 1842

\bibitem{PIE-Ca-1} Altun Z and Kelly H P 1985 {\it Phys. Rev.} A {\bf 31} 3711

\bibitem{PIE-Ca-2} Hansen J E and Scott P 1986 {\it Phys. Rev.} A {\bf 33} 3133

\bibitem{Salomonson} Salomonson S, Carter S L, and Kelly H P 1989 {\it Phys. Rev.} A {\bf 39} 5111

\bibitem{Burgdoerfer} Andersson L R and Burgd\"orfer J 1994 
{Nucl. Instrum. Meth. Phys. Res. B} {\bf 87} 167

\bibitem{Gorczyca} Gorczyca T W and Badnell N R 1997 J. Phys. B: At. Mol. Opt. Phys. {\bf 30} 3897

\bibitem{Greene} van der Hart H W, Meyer K W, and Greene C H 1998 {\it Phys. Rev. A} {\bf 57} 3641 

\bibitem{Keifets} Kheifets A S and Bray I 1998 {\it Phys. Rev. A} {\bf 57} 2590; \\
1998 {\it Phys. Rev. A} {\bf 58} 4501 

\bibitem{Bizau} Bizau J M and Wuilleumier F J 1995 
{\it J. Electron. Spectrosc. Relat. Phenom.} {\bf 71} 205

\bibitem{Wehlitz} Wehlitz R, Sellin I A, Hemmers O, Whitfield S B, Glans~P,
Wang H, Lindle D W, Langer B, Berrah N, Viefhaus J and Becker U 1997
{\it J. Phys. B: At. Mol. Opt. Phys.} {\bf 30} L51

\bibitem{2CPI} Najjari B, Voitkiv A~B and M\"{u}ller C 2010 
{\it Phys. Rev. Lett.} {\bf 105} 153002

\bibitem{Perina} Pe\v{r}ina J, Luk\v{s} A, Pe\v{r}inov{\'a} V and Leo{\'n}ski W 2011 {\it Phys. Rev. A} {\bf 83} 053416;\\ Pe\v{r}inov{\'a} V, Luk\v{s} A, K\v{r}epelka J and Pe\v{r}ina J 2014 {\it ibid.} \textbf{90} 033428

\bibitem{2CPI-coll}
Voitkiv A B, M\"uller C, Zhang S F and Ma X 2019 {\it New J. Phys.} {\bf 21} 103010

\bibitem{2CPI-mol}
Gr\"ull F, Voitkiv A~B and M\"uller C 2020 {\it Phys. Rev. A} {\bf 102} 012818;\\
2022 {\it J Phys B: At. Mol. Opt. Phys.} {\bf 55} 245101

\bibitem{ETI} A related process is decay via excitation-transfer ionization; see
Gokhberg K, Trofimov A B, Sommerfeld T and Cederbaum L S 2005 {\it Europhys. Lett.} {\bf 72} 228

\bibitem{2CPIexp}
Trinter F \textit{et al.} 2013 {\it Phys. Rev. Lett.} \textbf{111} 233004\\
Mhamdi A {\it et al.} 2018 {\it Phys. Rev.} A {\bf 97} 053407

\bibitem{Hergenhahn} Hans A, Schmidt P, Ozga C, Richter C, Otto H, Holzapfel~X, Hartmann G,
Ehresmann A, Hergenhahn~U and Knie A 2019 {\it J. Phys. Chem. Lett.} {\bf 10} 1078

\bibitem{Matthew} Matthew J and Komninos Y 1975 {\it Surf. Sci.} {\bf 53} 716

\bibitem{ICD} Cederbaum L~S, Zobeley J and Tarantelli F 1997 
{\it Phys. Rev. Lett.} \textbf{79} 4778

\bibitem{ICD-Review1}
Hergenhahn U 2011 {\it J. Electron Spectrosc. Relat. Phenom.} {\bf 184} 78

\bibitem{ICD-Review2} 
Jahnke T 2015 {\it J. Phys. B} {\bf 48} 082001

\bibitem{ICD-Review3}
Jahnke T, Hergenhahn U, Winter B, D\"orner R, Fr\"uhling~U, Demekhin P~V, Gokhberg K, 
Cederbaum~L~S, Ehresmann A, Knie A and Dreuw A 2020 {\it Chem. Rev.} {\bf 120} 20

\bibitem{SPDI-He2e-exp} 
Havermeier T {\it et al.} 2010 {\it Phys. Rev. Lett.} \textbf{104} 153401

\bibitem{SPDI-He2e-theor} 
Najjari B and Voitkiv A B 2021 {\it Phys. Rev. A} {\bf 104} 033104  

\bibitem{Hans-2023} Hans A {\it et al.} 2023 {\it Phys. Rev. Res.} {\bf 5} 013055

\bibitem{SPDI-ETMD-theo} Stumpf V, Kryzhevoi N~V, Gokhberg K and Cederbaum L~S 
2014 {\it Phys. Rev. Lett.} \textbf{112} 193001

\bibitem{SPDI-ETMD-exp} LaForge A~C {\it et al.} 2016 {\it Phys. Rev. Lett.} \textbf{116} 203001

\bibitem{DICD-theo} Averbukh V and Cederbaum L S 2006 {\it Phys. Rev. Lett.} \textbf{96} 053401

\bibitem{Fedyk} Fedyk J, Gokhberg K and Cederbaum L~S 2021 {\it Phys. Rev. A} {\bf 103} 022816

\bibitem{DICD-exp} LaForge A C, Shcherbinin M, Stienkemeier F, Richter R, Moshammer R, Pfeifer T and Mudrich M 2019 {\it Nature Phys.} \textbf{15} 247

\bibitem{Eckey2020} Eckey A, Voitkiv A B and M\"uller C 2020 
{\it J. Phys. B: At. Mol. Opt. Phys.} \textbf{53} 055001

\bibitem{NIST} Kramida A, Ralchenko Yu, Reader J and NIST ASD Team
2022 Atomic spectra database of the National Institute of
Standards and Technology (NIST) Gaithersburg version
5.10 (available at: https://physics.nist.gov/asd)

\bibitem{ICD-formula}
Gr\"ull F, Voitkiv A B and M\"uller C 2019 {\it Phys. Rev.} A {\bf 100} 032702

\bibitem{ICDexc} 
In such a case, the additional contribution to ICD mentioned below Eq.~\eqref{Gamma_ICD}, 
which leads to ionization-excitation of atom $A$, could compete with the usual ICD rate. 
It amounts to $\Gamma^{({\rm exc})}_{_{\rm ICD}}\approx\frac{3c^4}{2\pi\omega_{ge}^4 R^6}\,
\Gamma_{\rm rad}\,\sigma_{_{\rm PIE}}(\omega_{ge})$, for sufficiently large interatomic distances.

\bibitem{Sebastian} Steinh\"auser S 2018 B.Sc. thesis, Heinrich Heine University D\"usseldorf

\end{thebibliography}
\end{document}